\documentclass{jpp}
\usepackage{graphicx}

\usepackage[utf8]{inputenc}
\usepackage[T1]{fontenc}
\usepackage{amsmath}
\usepackage{color}
\usepackage{bm}
\usepackage[hidelinks]{hyperref}
\definecolor{doc}{rgb}{0.0,0.5,0.0}
\hypersetup{colorlinks=true,linkcolor=blue,citecolor=magenta}

\shorttitle{On the accuracy of binary collisions}
\shortauthor{T.P. Kiviniemi, E. Hirvijoki and A.J.Virtanen}

\title{On the accuracy of the binary-collision
algorithm in particle-in-cell simulations of magnetically confined fusion plasmas}

\author{Timo P. Kiviniemi\aff{1}
  \corresp{\email{timo.kiviniemi@aalto.fi}},
  Eero Hirvijoki\aff{1}
 \and Antti J. Virtanen\aff{1}}

\affiliation{\aff{1}Department of Applied Physics, Aalto University, P.O. Box 11100, 00076 AALTO, Finland}

\begin{document}

\maketitle

\begin{abstract}
    Ideally, binary collision algorithms conserve kinetic momentum and energy. In practice, the finite size of collision cells and the finite difference in the particle locations affect the conservation properties. In the present work, we investigate numerically how the accuracy of these algorithms is affected when the size of collision cells is large compared to gradient scale length of the background plasma, a parameter essential in full-$f$ fusion plasma simulations. Additionally, we discuss implications for the conserved quantities in drift-kinetic formulations when fluctuating magnetic and electric fields are present: we suggest how the accuracy of the algorithms could potentially be improved with minor modifications.
\end{abstract}


\section{Introduction}
Charged particles in a plasma interact with each other through the long-range Coulomb collisions and, in a particle-in-cell simulation, these interactions can be modelled with the so-called binary collision methods. The two widely-used schemes are by \citet{takizukaabe} and by \citet{nanbu}. If equal particle weights are used, both these methods preserve kinetic momentum and energy in local homogeneous simulations, which explains the popularity of these two schemes. Also, the convergence properties of the methods are well established. In \citet{wanglin}, collisional relaxation rates from these models are evaluated in a spatially homogeneous plasma with no electric field or magnetic fields and the  accuracy of the methods is compared as a function of time step and number of test particles per cell showing a $\mathcal{O}(\sqrt{\Delta t})$ dependency for the accuracy of electron-electron collisions while including electron-ion collisions was independent of $\Delta t$.
Nanbu's method was further tested by \citet{dimitswang} but again excluding fields. In global simulations, including configuration-space effects, the conservation properties generally depend on time step, number of test particles, particle sampling method, interpolation schemes and implementation of electromagnetic fields as well. The account of these effects is less established.

The importance of momentum and energy conservation itself depends on what quantity one is interested in, the time scale of the simulations (compared to collision time) and, also, the relative importance of the collisions compared to, e.g., turbulent effects and particle noise. In \citet{kivinieminf1999}, momentum-conserving binary-collision model and test-particle collision model were compared in case of externally induced radial field. Starting from zero parallel flow, it was shown that both methods give initially the same radial particle flux. After that the parallel velocity starts to develop in the momentum-conserving case and the flux decays. Since the development of parallel flow is a slower process than the changes in the mean radial electric field $\left<E_r\right>$, it is also possible to simulate a quasi-steady state of $E_r$ in order to investigate the accuracy of neoclassical analytic estimates as a function of gradient scale lengths, as done in \citet{kivinemicppnc}.
Violation of momentum conservation in numerical realizations can be mitigated, e.g., by forcing the curl of electric field $\bm{E}$ to zero with small adjustments in the radial component $E_r$ \citep{Heikkinen2012}. Finally, if also magnetic fluctuations are included in the simulation model, they contribute to both the conserved toroidal angular momentum and the energy (see, e.g., \citet{hirvijokiarxiv}). Consequently, the binary collision models should be considered in conjunction with the invariants of the collisionless dynamics.

In this work, we first take a look at the conventional conserved quantities and, as an example, demonstrate how even these can be inaccurate if the collision cell is too wide compared to the gradient scale length. After that we briefly discuss the conserved quantities in a drift-kinetic electromagnetic model, and propose how the accuracy of the conservation properties could potentially be improved while still using the standard binary collision model.

\section{Classic binary collisions model}
In performing particle-in-cell simulations and using the widely used binary collision models \citep{takizukaabe, nanbu}, 
collisional effects are naturally implemented so that they only change those parts of momentum and energy that directly depend on the particle distribution function. For example, in the 6D Vlasov-Maxwell model, the fields $\bm{E}$ and $\bm{B}$ are kept fixed during the collisional step. Correspondingly, the global functionals
\begin{align}
    P_F&=\sum_s \int m_s \bm{v}F_sd\bm{v}d\bm{x}\\
    E_F&=\sum_s \int \frac{1}{2}m_s |\bm{v}|^2F_sd\bm{v}d\bm{x},
\end{align}
should remain constant during the collisional step. Here, $m_s$ and $F_s$ are the mass and distribution function of species $s$, and  $\bm{x}$ and $\bm{v}$ are the location of particle in configuration and velocity space, respectively. 

In a binary collision algorithm with equal particle weights, implementing this strategy amounts to requesting that the kinetic energy and momentum are conserved in a pair-wise collision between the particles $p_1$ and $p_2$. Effectively, one requires that the following conditions are met
\begin{align}
    m_{p_1}\bm{v}_{p_1}(t_n)+m_{p_2}\bm{v}_{p_2}(t_n)&=m_{p_1}\bm{v}_{p_1}(t_{n+1})+m_{p_2}\bm{v}_{p_2}(t_{n+1}),\\
    m_{p_1}|\bm{v}_{p_1}(t_n)|^2+m_{p_2}|\bm{v}_{p_2}(t_n)|^2&=m_{p_1}|\bm{v}_{p_1}(t_{n+1})|^2+m_{p_2}|\bm{v}_{p_2}(t_{n+1})|^2,
\end{align}
where $t_{n+1} = t_n + \Delta t$ and $\Delta t$ is the time step. While convergence of these methods as a function of time step and number of test particles per cell has been demonstrated in homogeneous backgrounds \citep{wanglin}, we will next show that, even in the absence of fluctuating fields, these methods can be inaccurate if the background profiles change significantly within one collision cell.

\subsection{Effect of collision-cell width on the accuracy of binary collision models} 
\begin{figure}
\includegraphics[width=0.92\linewidth]{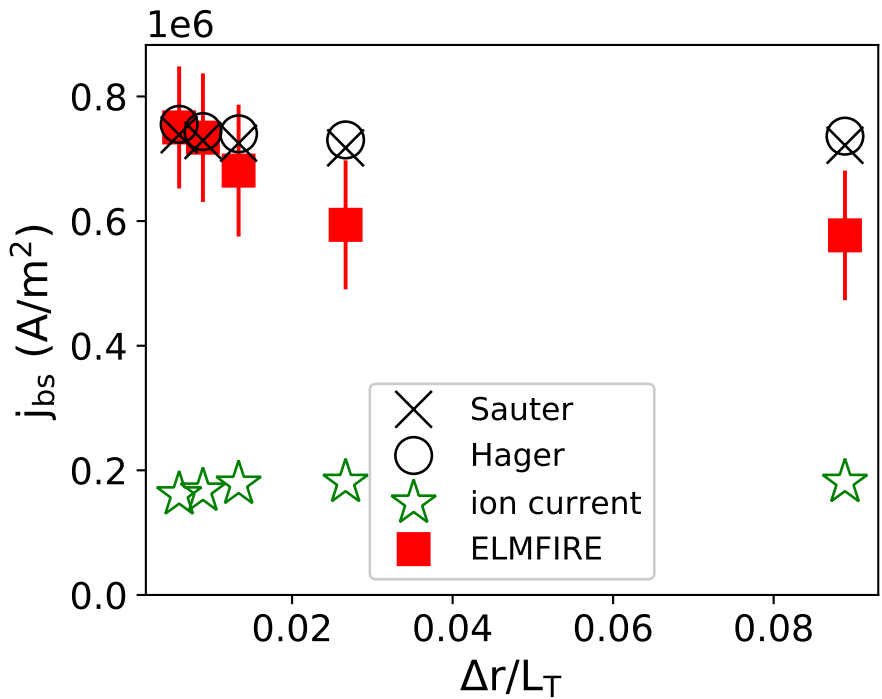} 

\caption{Scan of bootstrap current $j_{bs}$ as a function of normalized binary collision cell size, $\Delta r/L_T$. The $j_{bs}$ values are collected from the maximum current density location, and the error bars represent one standard deviation around the mean value and compared to analytic estimates of Sauter and Hager. The ion part of the total current is shown with stars. 
$\Delta r$ is the collision cell width in the radial direction and $L_T = 0.06$ m is the temperature  gradient scale length in the middle of the pedestal.}
\label{fig:slotscan}
\end{figure}

In axisymmetric tokamak geometry, the conservation of toroidal angular momentum is important to properly describe neoclassical and turbulent transport of particles and heat. In \citet{kivinemicppnc}, the effect of steep gradients on the accuracy of neoclassical analytic estimates was tested but not the effect of the size of the collision cell with respect to the gradient length scale of the background. Here, this effect is tested using the full-f particle-in-cell code
ELMFIRE \citep{korpilo2016}, running it in neoclassical mode and computing the bootstrap current similarly as in \citet{kiviniemippcf2014}. 
To investigate the effect of the number of collision cells on the
simulation accuracy and the numerical precision of the results, the number of cells, $N_{cells} = N  \times N$, was varied in a scan as N = 30, 100, 200, 300, 450 (see Fig. \ref{fig:slotscan}). The important relation in  such study is the relation between the cell width and the gradient scale lengths but since the bootstrap current itself heavily depends on the profiles we keep gradient scale lengths fixed as $L_n = L_T = 0.06$~m.
The simulation
domain was thus partitioned in N uniformly distributed collision cells in
both radial and poloidal directions. The total number of particles was kept
fixed between the simulations. For the case N = 450 the number of particles per
cell is approximately 300 on average, which is sufficiently high for convergence of the collisional rates \citep{wanglin}. 

The formation of bootstrap current is a two-step process where, at first, the ion current results from orbit topologies. As a second step, this ion current is transferred to electrons via collisions. Both of these steps can be affected by the cell width but, as seen in Fig.~\ref{fig:slotscan}, the ion current is small compared to total current and is not significantly affected by the number of cells. Thus, collisions are mostly responsible for the cell-width effect.
For the total current, increasing the number of binary-collision cells improved the quantitative agreement between the converged ELMFIRE simulation result and the analytical estimates of \citet{sauter} and \citet{hager}. With $300 \times 300$ collision cells, the simulated mean bootstrap current density is within 3\% from both theoretical predictions. Adding more collision cells did not notably change the result, but with fewer cells, ELMFIRE predicts distinctly lower $j_{bs}$ values. The two cases with the sparsest grids remain well below their corresponding analytical estimates, the relative difference for both is around 20 \%. The simulated $j_{bs}$ experiences strong temporal fluctuations which produces significant uncertainties. The error bars in Fig. \ref{fig:slotscan} illustrate one standard deviation from the mean and their size does not change when the number of collision cells is altered. 

Part of this puzzle is that the particles are paired for collisions from finite sized neighborhoods, the extent of which is determined by the number of collision cells used. The denser the grid is, the smaller the volume one cell covers. The binary
collision operator used in ELMFIRE assumes that the plasma background properties stay similar within each of these collision cells. Most importantly
the background density and temperature should not vary substantially over one collision cell as the collision frequency which determines the scattering angle in binary collision model
depends directly on background temperature through Coulomb logarithm
but also implicitly through the statistical increase in relative velocity as the average
velocities of particles are significantly different in the inner and outer side of the 
collision cell. In addition, it is also directly proportional to density.
 During a
simulation, the particle density is sampled to the simulation grid, and thus,
the density considered in the collisions is fixed in each of the spatial grid cells.
The smallest studied collision grid had size 
$30 \times 30$ which is sparser than the
$50 \times 50$ grid describing the density background in the radial and poloidal
directions, violating the assumption for the collision operator. For the larger
collision grids tested in the scan, the resolution for the density is no longer a
limiting factor, but accurate enough temperature resolution is also required.

The plasma temperature profile determines the speeds of individual particles. In particular, across a steep pedestal, the temperature changes abruptly
and so do the velocities of the particles in that region. When particles collide, 
their relative velocity is scattered and new velocities for both particles
are calculated from the result. The collision process inevitably introduces
non-locality in the updated particle velocities because the colliding particles
are paired at random within a collision cell. The random pairing is an approximation compared to true collisions but is often considered sufficient in
simulation. However, depending on the used grid size, the particles can have
significant distance between each other. Even if at the continuous level the Landau operator is local and describes Coulomb collisions at a resolution comparable to that of the Debye length---the effective distance beyond which the interaction is screened---practical implementations in kinetic simulations of fusion plasmas rarely are able to resolve this distance.

The differences observed between
the studied simulation cases could result, e.g., from the introduced finite spatial sampling of the background temperature. The thermal speed of the
particles' relate to temperature through the equation for
$v_{th} = \sqrt{2T /m}$, and
the change of average temperature within a collision cell with radial width $\Delta r$
can be approximated by $\Delta r \partial_r T = (\Delta r/L_T )T$ with $L_T$ the temperature gradient length scale. 
Inside a cell of the
sparsest studied collision grid, the difference in temperature 
$\Delta r/L_T$ can be
around 10\% when 
$L_T = 0.06$ m, which at the very high temperatures involved
becomes significant. In contrast, with a grid size $300 \times 300$, the difference is less than 1\%. Since the collision frequency is a function of the plasma temperature, one can expect a direct effect from not resolving the temperature accurately enough. Finally, numerical estimation of the bootstrap current requires accurate
modeling of parallel flow velocity and, in the ELMFIRE simulations, the
average parallel velocity of a particle species is sampled from the individual
particle velocities which allows small inaccuracies to accumulate. Therefore,
more accurate description of the velocity distribution obtained with denser
collision grids is likely improving the simulated bootsrap current $j_{bs}$.

\section{Conserved quantities in an electromagnetic drift-kinetic model}
If electromagnetic fluctuations are included in the simulations, they affect the quantities that are conserved by the collisionless dynamics. Considering then also the collisional dynamics, the binary collision model should retain the invariants of the colllision-free model. For fusion plasmas, an electromagnetic drift-kinetic model that results as the $k_\perp \rho\ll 1$ limiting case of the electromagnetic gyrokinetic model \citep{Burby_2019} is of particular interest. The analysis of the conserved quantities for such a model can be found, e.g., in \citet{hirvijokiarxiv}.

For this case, the "kinetic-momentum"- and the "kinetic energy"-like functionals, that the binary-collision algorithm should leave invariant for fixed values of the fields $\bm{E}_1$ and $\bm{B}_1$, are given by 
\begin{align}
    P_F&=\sum_s\int F_s \left(e_s\bm{A}_0+m_su\bm{b}_0 -\frac{\partial K_s}{\partial \bm{E}_1}\times\bm{B}_1\right)\cdot\bm{e}_{\varphi}du d\mu d\bm{x},\\
    E_F&=\sum_s\int \Big(K_s-\frac{\partial K_s}{\partial \bm{E}_1}\cdot\bm{E}_1\Big)F_{s}du d\mu d\bm{x},
\end{align}
where the summation over $s$ again refers to particle species.
Here, $\bm{B}_0=\nabla\times\bm{A}_0$ is the background magnetic field, with $\bm{b}_0=\bm{B}_0/|\bm{B}_0|$ the corresponding unit vector. The dynamical fields in the system are the distributional densities $F_s$, which include the phase-space Jacobian, and the electric and magnetic field perturbations $\bm{E}_1$ and $\bm{B}_1$. The single drift-center kinetic energy function in the model (\citet{hirvijokiarxiv}) is given by
\begin{align}
    K &= \frac{1}{2}mu^2+\mu |\bm{B}_0|\left(1+\frac{\bm{b}_0\cdot\bm{B}_1}{|\bm{B}_0|}+\frac{|\bm{B}_{1\perp}|^2}{2|\bm{B}_0|^2}\right)-\frac{m}{2|\bm{B}_0|^2}|\bm{E}_{1\perp}+u\bm{b}_0\times\bm{B}_1|^2.
\end{align}

From the global functionals, we identify the individual particle contributions, namely 
\begin{align}
    P(\bm{x},u)&=\left(e\bm{A}_0+mu\bm{b}_0 -\frac{\partial K}{\partial \bm{E}_1}\times\bm{B}_1\right)\cdot\bm{e}_{\varphi},\\
    E(\bm{x},u,\mu)&=K-\frac{\partial K}{\partial \bm{E}_1}\cdot\bm{E}_1.
\end{align}
Regardless of what exactly a conservative binary collision algorithm does, it should satisfy the pair-wise conservation of toroidal angular momentum and total energy
\begin{align}
    P_{1,n}+P_{2,n}&=P_{1,n+1}+P_{2,n+1},\\
    E_{1,n}+E_{2,n}&=E_{1,n+1}+E_{2,n+1},
\end{align}
with the notation $P_{1,n}\equiv P(\bm{x}_{1,t_n},u_{1,t_n})$ etc. and 
$(x_{t_n},u_{t_n},\mu_{t_n})$ and $(x_{t_{n+1}},u_{t_{n+1}},\mu_{t_{n+1}})$ referring to the particle coordinates before and after the collisional time step $\Delta t$.

The standard binary-collision algorithms, however, are not designed to preserve these particular invariants in the presence of the perturbations $\bm{E}_1$ and $\bm{B}_1$, resulting in deviations $\Delta P$ and $\Delta E$ such that 
\begin{align}
    P_{1,n}+P_{2,n}&=P_{1,n+1}+P_{2,n+1}+\Delta P,\\
    E_{1,n}+E_{2,n}&=E_{1,n+1}+E_{2,n+1}+\Delta E.
\end{align}

Since the field fluctuations by definition are supposed to be small, the new values for velocities from a standard binary collision step nevertheless are expected to approximately retain the invariants, and significant errors to accumulate only over time. Consequently, a small perturbation, e.g., a shift in the location or velocity of particle one, $\bm{x}_1$, at every time step, could potentially be used to make the deviations $\Delta E$ and $\Delta P$ to vanish.

\subsection{Potential corrections to conserving P and E}

\begin{figure}
\includegraphics[width=0.96\linewidth]{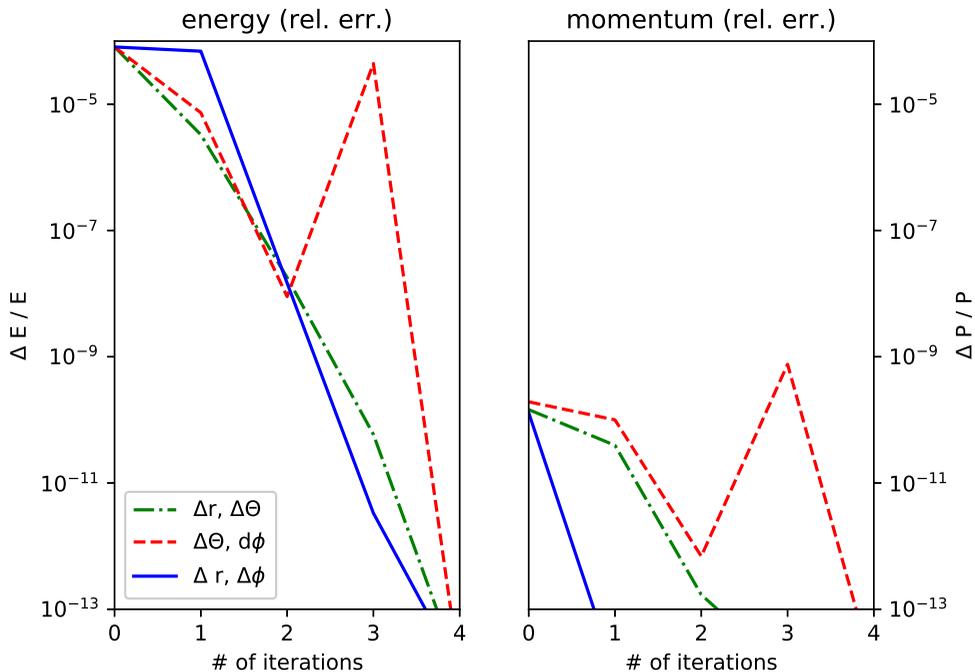}
\caption{Relative change of E and P just after binary collisions
(index "0" in xlabel) and after iterative corrections.}
\label{fig:momene}
\end{figure}
Using $(x_1,x_2,x_3)$ for the configuration space coordinates of particle 1 after the standard binary collision step has been taken and the errors $\Delta P$ and $\Delta E$ are known, we could adjust, say, two of the coordinates according to
\begin{equation}
\begin{bmatrix}
\Delta x_1  \\
\Delta x_2
\end{bmatrix}
=
\begin{bmatrix}
dP/d{x_1} &  dP/d{x_2}\\
dE/d{x_1} & dE/d{x_2}
\end{bmatrix}^{-1}
\begin{bmatrix}
\Delta P \\
\Delta E
\end{bmatrix}\label{cormat}
\end{equation}
to reduce the error. Further, this corrective step can be iterated to suppress the error significantly.
In three-dimensional case, there is freedom to choose any two out of the three available components for tuning the quantities $P$ and $E$. In toroidal coordinates, the relative errors in momentum appear to be quite small. The correction terms depend much on the numerical parameters and mainly on radial coordinate, $P \approx P(r)$. Other corrections are very small.

In Fig.~\ref{fig:momene}, the correction method is tested with a simple test case for sinusoidal $|\bm{B}_1|/|\bm{B}_0| = \mathcal{O}(10^{-3})$ fluctuations. Repeated binary
collisions of two particles are carried out and, after each binary collision,
$P$ and $E$ are corrected using Eq.~\eqref{cormat}
("0" refers to error just after BC). The standard deviation of the error compared to $P$ ($E$) before the binary collision is shown.
It can be seen that correction in $P$ is small, $\mathcal{O}(10^{-10})$, while relative error in $E$ is order of $10^{-4}$.
Tuning with $\Delta r$ together with poloidal ($\Delta \theta$) or toroidal ($\Delta \phi$) correction shows the best performance confirming that in practise the radial coordinate $r$ tunes $P\approx P(r)$ after which the fine tuning of E is done with either $\theta$ or $\phi$.

If a scheme, such as the one described above, is adopted to enforce the conservation properties, one can expect at least some level of artificial transport. We can try to estimate the level of such induced transport in the following manner. Say the bare binary collision algorithm, without fluctuating fields, provides a change in the parallel velocity $\Delta u$. In the presence of fluctuations, this induces an additional change in the toroidal canonical momentum which we can approximate from 
\begin{align}
    \Delta P\sim m\Delta u \frac{\partial}{\partial u}\frac{\partial K}{\partial \bm{E}_1}\times\bm{B}_1\cdot\bm{e}_{\phi}&\sim m \Delta u\left(\frac{B_1}{B_0}\right)^2 R.
\end{align}
If we shift the particle position in radial direction, the dominant change in the canonical toroidal momentum becomes
\begin{align}
    \Delta P\sim e\Delta r \frac{\partial \Psi_p}{\partial r}\sim e\Delta rRB_{0,p}
\end{align}
where $B_{0,p}$ is the poloidal component of the unperturbed magnetic field. In trying to counter the change in the toroidal momentum credited for the fluctuating fields during a Coulomb collision, the particle's radial position then needs to be shifted by the amount
\begin{align}
    \Delta r\sim\frac{m}{eB_p}\left(\frac{B_1}{B_0}\right)^2\Delta u.
\end{align}
The associated diffusion coefficient can be estimated from $D\sim(\Delta r)^2/\Delta t$, which together with $(\Delta u)^2/\Delta t\sim \nu u^2$ and $\nu$ denoting the collision frequency, leads to the estimate
\begin{align}
    D=\frac{(\Delta r)^2}{\Delta t}\sim \left(\frac{B_0}{B_{0,p}}\right)^2\left(\frac{B_1}{B_0}\right)^4\rho_0^2\nu.
\end{align}
Even if the magnetic fluctuations were comparable to the poloidal magnetic field, the term $(B_0/B_{0,p})^2(B_1/B_0)^4$ would remain considerably less than one. Consequently, we expect that the transport from the corrective algorithm would remain at most at the level of classical diffusion and likely be significantly less than that.

\section{Conclusions}
In this work, we have demonstrated that the accuracy of the widely used binary collision algorithm decreases when the collision grid cell size reaches a significant fraction of the gradient scale length of the plasma background. This indicates that, while the standard binary collision algorithm works well in homogeneous backgrounds, either very small collision cell sizes should be used in the steep gradient regions at the tokamak edge or the collision algorithm modified. We then suggested one possibility to modify the existing binary collision algorithms to regain the conservation of the quantities important in transport simulations when electromagnetic fluctuations are present. We expect such minor modifications to be also practical enough for implementations.

\section{Acknowledgements}

The work has been supported by the Academy of Finland (T.K., grant number 316088), (E.H., grant number 315278) and is part (T.K.) of
Eurofusion enabling research projects "Model for reactor relevant pedestals" (CCFE-04)
and "MAGYK: Mathematics and Algorithms for GYrokinetic and Kinetic models" (MPG-04). This work has been carried out within the framework of the EUROfusion Consortium and has received funding from the Euratom research and training programme 2014-2018 and 2019-2020 under grant agreement No 633053. The views and opinions expressed herein do not necessarily reflect those of the European Commission. 
CSC – IT Center for Science is acknowledged for generous allocation of computational resources for this work.

\section{Declaration of interests}

The authors report no conflict of interest.

\bibliographystyle{jpp}
\bibliography{references}

\end{document}